\documentclass[12pt]{article}
\usepackage{times}
\usepackage{geometry}
\usepackage[utf8]{inputenc}
\usepackage{enumitem,amssymb}
\usepackage{ragged2e}
\usepackage{graphicx}
\usepackage{fancyhdr}
\usepackage[
    natbib=true,
    style=numeric,
    sorting=none
]{biblatex}
\newlist{thematic}{itemize}{8}
\setlist[thematic]{label=$\square$}
\usepackage{pifont}

\begin{document}
\raggedright
\huge{
Prioritizing High-Precision Photometric Monitoring of Exoplanet and Brown Dwarf Companions with JWST}
\linebreak
\normalsize
Ben J. Sutlieff$^{1}$, Xueqing Chen$^{1}$, Pengyu Liu$^{1}$, Emma E. Bubb$^{1}$, Stanimir A. Metchev$^{2}$, Brendan P. Bowler$^{3}$, Johanna M. Vos$^{4}$, Raquel A. Martinez$^{5}$, Genaro Suárez$^{6}$, Yifan Zhou$^{7}$, Samuel M. Factor$^{3}$, Zhoujian Zhang$^{8}$, Emily L. Rickman$^{9}$, Arthur D. Adams$^{10}$, Elena Manjavacas$^{9}$, Julien H. Girard$^{11}$, Bokyoung Kim$^{1}$, Trent J. Dupuy$^{1}$ | \textit{(Affiliations can be found after the references)}

\large


\noindent \textbf{Thematic Areas (Check all that apply):} \linebreak \linebreak $\boxtimes$ (Theme A) Key science themes that should be prioritized for future JWST and HST observations 
\linebreak $\boxtimes$ (Theme B) Advice on optimal timing for substantive follow-up observations and mechanisms for enabling exoplanet science with HST and/or JWST \linebreak
$\square$ (Theme C) The appropriate scale of resources likely required to support exoplanet science with HST and/or JWST 
 \linebreak
$\square$ (Theme D) A specific concept for a large-scale ($\sim$500 hours) Director’s Discretionary exoplanet program to start implementation by JWST Cycle 3.
 \linebreak

\justify{
\textbf{Summary:} We advocate for the prioritization of high-precision photometric monitoring of exoplanet and brown dwarf companions to detect brightness variability arising from features in their atmospheres. Measurements of photometric variability provide not only an insight into the physical appearances of these companions, but are also a direct probe of their atmospheric structures and dynamics, and yield valuable estimates of their rotation periods. JWST is uniquely capable of monitoring faint exoplanet companions over their full rotation periods, thanks to its inherent stability and powerful high-contrast coronagraphic imaging modes. Rotation period measurements can be further combined with measurements of $v \sin i$ obtained using high-resolution spectroscopy to infer the viewing angle of a companion. Photometric monitoring over multiple rotation periods and at multiple epochs will allow both short- and long-term time evolution in variability signals to be traced. Furthermore, the differences between the layers in a companion's atmosphere can be probed by obtaining simultaneous photometric monitoring at different wavelengths through NIRCam dual-band coronagraphy. Overall, JWST will reach the highest sensitivities to variability to date and enable the light curves of substellar companions to be characterised with unprecedented cadence and precision at the sub-percent level.

}

\pagebreak
\justify{

\textbf{Anticipated Science Objectives:}
Time-resolved photometric monitoring has revealed periodic variations in the light curves of brown dwarfs and exoplanet analogues, with young objects close to the L/T transition showing the highest variability rates (see Fig. 1) \cite{2015ApJ...799..154M, 2022ApJ...924...68V}. Most young, giant, directly-imaged exoplanets possess similar temperatures and surface gravities to L/T dwarfs, and thus are expected to show similarly high levels of variability. However, the high-contrasts and close angular separations of these companions have made it challenging to detect them at a sufficiently high signal-to-noise to measure their variability. Furthermore, while substellar objects are fast rotators (typically 3-20~h, \cite{2006ApJ...647.1405Z}), ground-based observations can rarely cover their complete rotation periods. JWST is the only facility capable of high-precision photometric monitoring of exoplanet companions over their full periods. Positive detections of variability in the light curves of companions observed with JWST will yield estimates of their periods, and simultaneous monitoring at different wavelengths will allow us to probe different layers of their atmospheres and hence gain new insights into their 3D structure (Fig. 2).

 \vspace{0.2cm}

\textbf{Urgency}: Initial photometric monitoring observations should ideally be performed early in JWST's lifetime to allow additional observations of variable companions in future cycles. This will enable the characterisation of long-term variability trends arising from evolving weather systems or cloud bands which rotate within the companion's atmosphere and produce waves on a global scale \cite{2017Sci...357..683A, 2021MNRAS.502..678T}.

\textbf{Risk/Feasibility}: The level of photometric stability as a function of time that can be achieved using JWST's coronagraphic modes has yet to be empirically measured, so the variability amplitudes that can be reached with this technique remain uncertain. However, similar space-based variability observations with HST and Spitzer have proven highly successful, reaching sensitivities to variability at the $\sim$0.1-1\% level \cite[e.g.][]{2015ApJ...799..154M, 2022ApJ...924...68V, 2013ApJ...768..121A, 2016ApJ...818..176Z}. Moreover, non-detections of variability place valuable constraints on variability occurrence rates and the atmospheric properties of companions.

\textbf{Timeliness}: Variability measurements of exoplanetary companions will inform observations with the upcoming Extremely Large Telescopes (ELTs), the large mirror diameters of which will enable photometric monitoring of similar companions at closer angular separations.

\textbf{Cannot be accomplished in the normal GO cycle}:  Long-term variability trends can only be detected through photometric monitoring observations obtained at multiple epochs. Efforts to characterise variable targets would therefore benefit from an observing strategy covering several cycles.

}

\pagebreak

\begin{figure*}[htpb]
   \centering
   \includegraphics[width=0.4\textwidth,clip,trim=0mm 0mm 0mm 0mm,angle=0]{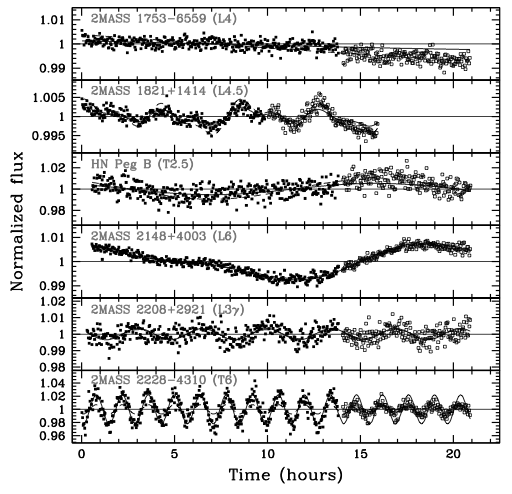}
\includegraphics[width=0.59\textwidth,clip,trim=0mm 0mm 0mm 0mm,angle=0]{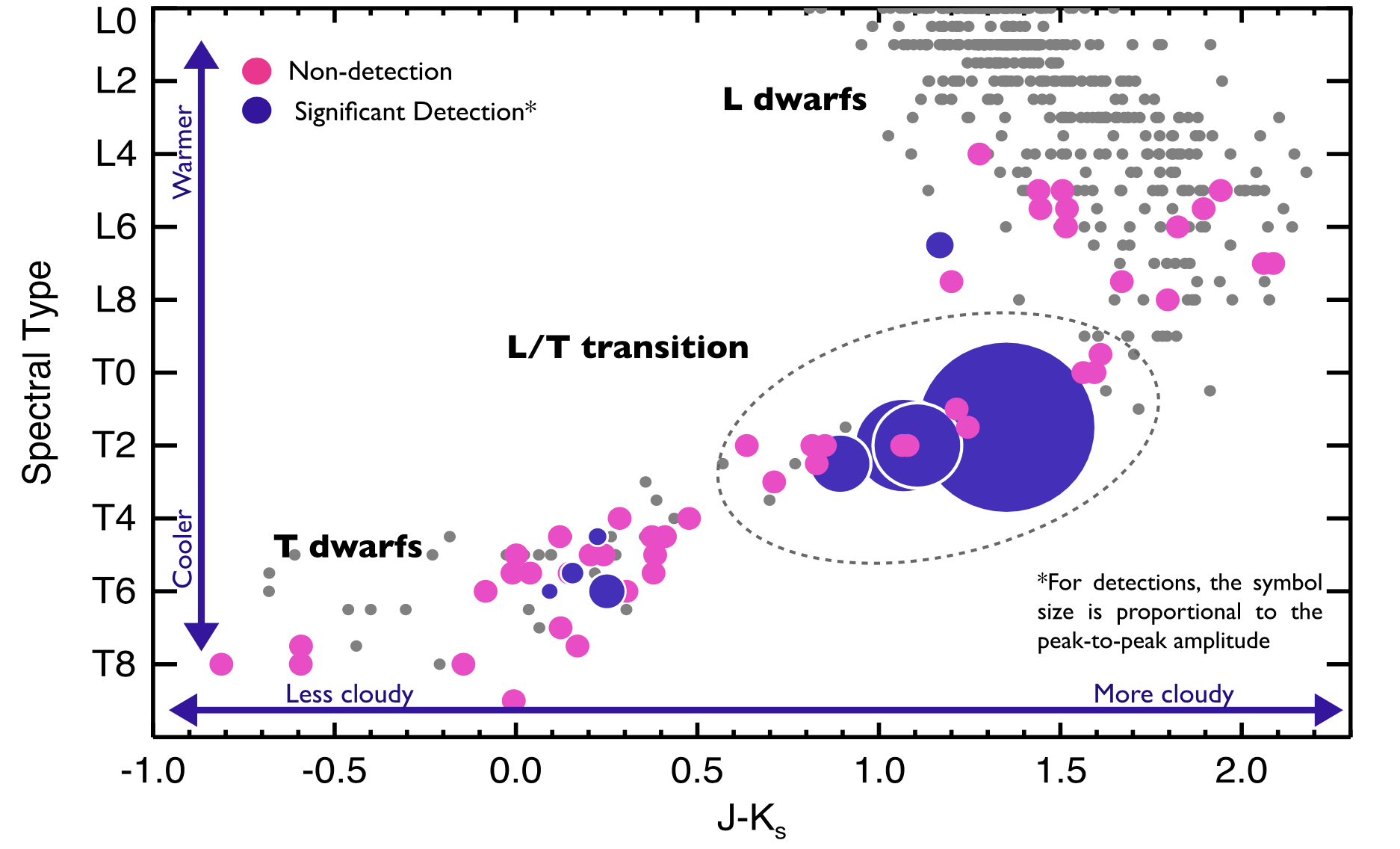}   \\  \caption{\textbf{Left:} Variability light curves for different classes of brown dwarf observed with Spitzer. Many objects vary at the $\ge1\%$ level. \textbf{Right:} Variability detections for brown dwarfs of different spectral classes. Objects close to the L/T transition show the highest variability rates and amplitudes. Directly imaged exoplanets have similar temperatures and low surface gravities to these objects, and so are expected to be similarly highly variable. Figures adapted from \cite{2015ApJ...799..154M, 2014ApJ...793...75R}.} 
\end{figure*}

\begin{figure*}[htpb]
   \centering
   \includegraphics[scale=0.275,clip,trim=0mm 0mm 0mm 0mm,angle=0]{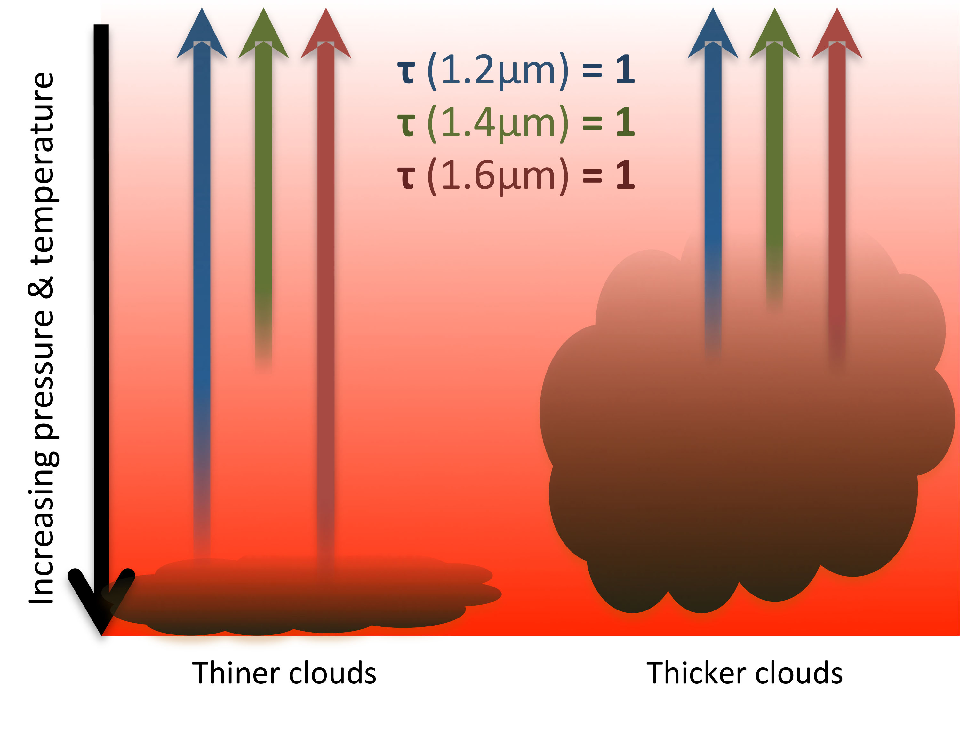}\\  \caption{Different wavelengths probe different layers of an exoplanet's atmosphere. Hence, simultaneous monitoring in different bands can provide valuable information about their 3D structure. Figure adapted from \cite{2018haex.bookE..94A}.} 
\end{figure*}
\pagebreak
\printbibliography

@ARTICLE{2022ApJ...924...68V,
       author = {{Vos}, Johanna M. and {Faherty}, Jacqueline K. and {Gagn{\'e}}, Jonathan and {Marley}, Mark and {Metchev}, Stanimir and {Gizis}, John and {Rice}, Emily L. and {Cruz}, Kelle},
        title = "{Let the Great World Spin: Revealing the Stormy, Turbulent Nature of Young Giant Exoplanet Analogs with the Spitzer Space Telescope}",
      journal = {\apj},
         year = 2022,
        month = jan,
       volume = {924},
       number = {2},
          eid = {68},
        pages = {68},
          doi = {10.3847/1538-4357/ac4502},
archivePrefix = {arXiv},
       eprint = {2201.04711},
 primaryClass = {astro-ph.EP},
       adsurl = {https://ui.adsabs.harvard.edu/abs/2022ApJ...924...68V}
}

@ARTICLE{2015ApJ...799..154M,
       author = {{Metchev}, Stanimir A. and {Heinze}, Aren and {Apai}, D{\'a}niel and {Flateau}, Davin and {Radigan}, Jacqueline and {Burgasser}, Adam and {Marley}, Mark S. and {Artigau}, {\'E}tienne and {Plavchan}, Peter and {Goldman}, Bertrand},
        title = "{Weather on Other Worlds. II. Survey Results: Spots are Ubiquitous on L and T Dwarfs}",
      journal = {\apj},
         year = 2015,
        month = feb,
       volume = {799},
       number = {2},
          eid = {154},
        pages = {154},
          doi = {10.1088/0004-637X/799/2/154},
archivePrefix = {arXiv},
       eprint = {1411.3051},
 primaryClass = {astro-ph.SR},
       adsurl = {https://ui.adsabs.harvard.edu/abs/2015ApJ...799..154M}
}

@ARTICLE{2016ApJ...818..176Z,
       author = {{Zhou}, Yifan and {Apai}, D{\'a}niel and {Schneider}, Glenn H. and {Marley}, Mark S. and {Showman}, Adam P.},
        title = "{Discovery of Rotational Modulations in the Planetary-mass Companion 2M1207b: Intermediate Rotation Period and Heterogeneous Clouds in a Low Gravity Atmosphere}",
      journal = {\apj},
         year = 2016,
        month = feb,
       volume = {818},
       number = {2},
          eid = {176},
        pages = {176},
          doi = {10.3847/0004-637X/818/2/176},
archivePrefix = {arXiv},
       eprint = {1512.02706},
 primaryClass = {astro-ph.EP},
       adsurl = {https://ui.adsabs.harvard.edu/abs/2016ApJ...818..176Z}
}

@ARTICLE{2013ApJ...768..121A,
       author = {{Apai}, D{\'a}niel and {Radigan}, Jacqueline and {Buenzli}, Esther and {Burrows}, Adam and {Reid}, Iain Neill and {Jayawardhana}, Ray},
        title = "{HST Spectral Mapping of L/T Transition Brown Dwarfs Reveals Cloud Thickness Variations}",
      journal = {\apj},
         year = 2013,
        month = may,
       volume = {768},
       number = {2},
          eid = {121},
        pages = {121},
          doi = {10.1088/0004-637X/768/2/121},
archivePrefix = {arXiv},
       eprint = {1303.4151},
 primaryClass = {astro-ph.EP},
       adsurl = {https://ui.adsabs.harvard.edu/abs/2013ApJ...768..121A}
}

@ARTICLE{2021MNRAS.502..678T,
       author = {{Tan}, Xianyu and {Showman}, Adam P.},
        title = "{Atmospheric circulation of brown dwarfs and directly imaged exoplanets driven by cloud radiative feedback: effects of rotation}",
      journal = {\mnras},
         year = 2021,
        month = mar,
       volume = {502},
       number = {1},
        pages = {678-699},
          doi = {10.1093/mnras/stab060},
archivePrefix = {arXiv},
       eprint = {2005.12152},
 primaryClass = {astro-ph.EP},
       adsurl = {https://ui.adsabs.harvard.edu/abs/2021MNRAS.502..678T}
}

@ARTICLE{2017Sci...357..683A,
       author = {{Apai}, D. and {Karalidi}, T. and {Marley}, M.~S. and {Yang}, H. and {Flateau}, D. and {Metchev}, S. and {Cowan}, N.~B. and {Buenzli}, E. and {Burgasser}, A.~J. and {Radigan}, J. and {Artigau}, E. and {Lowrance}, P.},
        title = "{Zones, spots, and planetary-scale waves beating in brown dwarf atmospheres}",
      journal = {Science},
         year = 2017,
        month = aug,
       volume = {357},
       number = {6352},
        pages = {683-687},
          doi = {10.1126/science.aam9848},
       adsurl = {https://ui.adsabs.harvard.edu/abs/2017Sci...357..683A}
}

@ARTICLE{2006ApJ...647.1405Z,
       author = {{Zapatero Osorio}, M.~R. and {Mart{\'\i}n}, E.~L. and {Bouy}, H. and {Tata}, R. and {Deshpande}, R. and {Wainscoat}, R.~J.},
        title = "{Spectroscopic Rotational Velocities of Brown Dwarfs}",
      journal = {\apj},
         year = 2006,
        month = aug,
       volume = {647},
       number = {2},
        pages = {1405-1412},
          doi = {10.1086/505484},
archivePrefix = {arXiv},
       eprint = {astro-ph/0603194},
 primaryClass = {astro-ph},
       adsurl = {https://ui.adsabs.harvard.edu/abs/2006ApJ...647.1405Z}
}

@INCOLLECTION{2018haex.bookE..94A,
       author = {{Artigau}, {\'E}tienne},
        title = "{Variability of Brown Dwarfs}",
    booktitle = {Handbook of Exoplanets},
         year = 2018,
       editor = {{Deeg}, Hans J. and {Belmonte}, Juan Antonio},
          eid = {94},
        pages = {94},
          doi = {10.1007/978-3-319-55333-7_94},
       adsurl = {https://ui.adsabs.harvard.edu/abs/2018haex.bookE..94A}
}

@ARTICLE{2014ApJ...793...75R,
       author = {{Radigan}, Jacqueline and {Lafreni{\`e}re}, David and {Jayawardhana}, Ray and {Artigau}, Etienne},
        title = "{Strong Brightness Variations Signal Cloudy-to-clear Transition of Brown Dwarfs}",
      journal = {\apj},
         year = 2014,
        month = oct,
       volume = {793},
       number = {2},
          eid = {75},
        pages = {75},
          doi = {10.1088/0004-637X/793/2/75},
archivePrefix = {arXiv},
       eprint = {1404.3247},
 primaryClass = {astro-ph.SR},
       adsurl = {https://ui.adsabs.harvard.edu/abs/2014ApJ...793...75R}
}

\footnotesize
\noindent$^{1}$Institute for Astronomy, University of Edinburgh, Royal Observatory, Blackford Hill, Edinburgh, EH9 3HJ, UK\\
$^{2}$The University of Western Ontario, 1151 Richmond Avenue, London, ON N6A 3K7, Canada\\
$^{3}$Department of Astronomy, The University of Texas at Austin, Austin, TX 78712, USA\\
$^{4}$School of Physics, Trinity College Dublin, The University of Dublin, Dublin 2, Ireland\\
$^{5}$Department of Physics \& Astronomy, University of California, Irvine, Irvine, CA 92697, USA\\
$^{6}$Department of Astrophysics, American Museum of Natural History, New York, NY 10024, USA\\
$^{7}$Department of Astronomy, University of Virginia, Charlottesville, VA 22904, USA\\
$^{8}$Department of Astronomy \& Astrophysics, University of California, Santa Cruz, Santa Cruz, CA 95060, USA\\
$^{9}$European Space Agency (ESA), ESA Office, Space Telescope Science Institute, 3700 San Martin Drive, Baltimore, MD 21218, USA\\
$^{10}$Department of Earth and Planetary Sciences, University of California, Riverside, 900 University Ave, Riverside, CA~92521, USA\\
$^{11}$Space Telescope Science Institute, 3700 San Martin Drive, Baltimore, MD 21231, USA

\end{document}